\documentclass{aastex631}

\usepackage{xspace}
\newcommand{\mc}{$\mathbf{MC}$\xspace}
\newcommand{\mI}{$\mathbf{M}_1$\xspace}

\begin{document}

\title{X-ray Spectra from General Relativistic RMHD Simulations of Thin Disks}

\author[0000-0002-6485-2259]{Nathaniel Roth}
\affiliation{Lawrence Livermore National Laboratory, P.O. Box 808, Livermore, CA 94550, USA}

\author{Peter Anninos}
\affiliation{Lawrence Livermore National Laboratory, P.O. Box 808, Livermore, CA 94550, USA}

\author[0000-0002-5786-186X]{P. Chris Fragile}
\affiliation{Department of Physics and Astronomy, College of Charleston, 66 George Street, Charleston, SC 29424, USA}

\author{Derrick Pickrel}
\affiliation{Department of Physics, University of California, 5200 North Lake Road, Merced, CA 95343, USA}

\begin{abstract}
We compare X-ray emission from several general relativistic, multi-frequency, radiation magnetohydrodynamic
simulations of thin black hole accretion disks with different accretion rates and spins. The simulations were performed
using the \mI closure scheme, resolved with twelve frequency (energy) bins logarithmically spaced 
from $5\times 10^{-3}$ to $5\times 10^3$ keV. We apply a general relativistic Monte Carlo transport
code to post-process the simulation data with greater fidelity in frequency resolution and Compton scattering treatment.
Despite the relatively few energy bins and Kompaneets approximation to Compton scattering utilized in the \mI
method, we find generally good agreement between the methods. Both produce prominent thermal profiles with peaks around
2 - 2.5 keV, where agreement is particularly strong and representative of the soft state. Both also find weaker
(lower luminosity) thermally sourced emission extending out to 100 keV due to the hotter innermost regions of the disks.
Inverse Compton scattering becomes increasingly effective at hardening spectral outputs with increasing black hole spin,
and becomes the dominant mechanism for photons that escape with energies between 10 to several hundred keV. 
At very high rates of spin the radiation flux in this upscattered component becomes comparable to the 
thermal flux, a phenomenon typically associated with intermediate states. 
Beyond $10^4$ keV, we observe faint, free-free emission from hot, optically thin coronal regions developing near the
horizon, common to both spinning and nonspinning black holes.
\end{abstract}

\keywords{Accretion (14) --- Radiative magnetohydrodynamics (2009) --- Relativistic disks (1388) --- Rotating black holes (1406) --- Low-mass X-ray binary stars (939)}

\section{Introduction} \label{sec:intro}

Despite decades of progress with numerical simulations of black hole accretion disks \citep[c.f.,][]{Abramowicz13}, it has only been fairly recently that serious attempts have been made to directly compare the results with observations. One reason was that early simulations did not include radiation; therefore, none of the quantities that could be extracted from the simulations had direct observational corollaries. This led groups to develop post-processing codes that could ``paint'' radiation onto the simulations after the fact and, thus, allow comparisons with observations \citep{Dexter09, Dolence09, Vincent11, Schnittman13, Kinch19, Kinch21, Kawashima23}. This method is particularly effective in low-luminosity systems where the radiation is weak enough to have limited effects on the dynamics. For example, this is what has been used to match most of the Event Horizon Telescope observations of M87 \citep{EHT19} and Sgr A* \citep{EHT22}.

Treatment of higher luminosity systems, where the radiation plays a dynamically important role, had to await the development of radiation magnetohydrodynamic (RMHD) schemes capable of performing global simulations \citep{Jiang14, McKinney14, Sadowski14, Fragile14} so that the radiation could be evolved along with the magnetohydrodynamic fields. Because the radiation is tracked and evolved throughout such simulations, these schemes made it possible to extract observables, such as light curves, directly from the simulations without extensive post-processing. However, the early RMHD simulations all used grey (frequency-integrated) opacities, so spectral information was still lacking and again had to be recovered via postprocessing \citep[e.g.,][]{Narayan16, Mills24}.

Spectral information is crucial, though, to the study, and likely the evolution, of black hole accretion systems. Spectra are used for estimating the mass \citep[e.g.,]
[]{Dotani97} and spin \citep{Davis06, Shafee06} of black holes. Spectral timing is also used to study the geometry of black hole accretion flows \citep{Uttley14}. More pertinently, the spectra of most black hole accretion systems are composed of multiple (soft and hard) components that likely originate from different physical regions of the accretion flow \citep[see][]{Done07}. Due to both geometric and strong curvature effects, photons from these different components have a good chance of intersecting other parts of the flow, which can then dramatically affect the overall thermodynamics if the intersecting photons are absorbed, scattered or otherwise reprocessed. 
In such scenarios, it may be necessary to retain spectral information about the radiation in order to properly capture its effect on the intercepted plasma, and hence the thermodynamics and structure of the accretion flow. Additionally, we anticipate that there may be many accreting systems, such as tilted or warped disks or tidal disruption events, where the reprocessing of radiation across a wide spectral range may be crucial to properly capturing the overall evolution. In such cases, gray-opacity evolution followed by spectral postprocessing is unlikely to be sufficient.

In light of this, groups including ours have recently begun to develop codes that can self-consistently and simultaneously evolve the radiative transport equations in multiple frequency bins in order to retain and exploit more of the spectral information within the simulations themselves \citep{Anninos20, Jiang22, Cheong23}. The focus of the current paper is to compare the spectral outputs from numerical simulations (described in the next section) performed using the integrated multi-frequency \mI method of \citet{Anninos20} with our own Monte Carlo radiation transport postprocessing code \citep{Roth22}. The goal is to confirm that the inline spectral treatment is reasonable, considering the limited resources available for 3D models that impose hard limits on spectral fidelity, while also expounding on the nature and source of the spectral outputs through the flexibility of postprocessing.

\section{Base Simulations} \label{sec:simulations}

In a previous paper \citep{Fragile23}, we presented several three-dimensional general relativistic, multi-frequency, 
RMHD simulations of radiatively efficient, thin black hole accretion disks with solar composition, and computed their
X-ray spectral emissions utilizing the \mI radiation closure scheme. 
These simulations covered a range of mass accretion rates and black hole spins, starting from a standard Novikov-Thorne
configuration \citep{Novikov73} seeded with a weak quadrupole magnetic field and evolved with the multi-group radiation transport capabilities of Cosmos++ \citep{Anninos20}. 
All calculations were performed on a 3-level, statically and logarithmically refined spherical-polar grid with increased resolution  near the black
hole and along the midplane of the disk. With a couple exceptions, the grid covered from just inside the innermost stable circular orbit (ISCO) radius 
to $160 r_G$, where $r_G=GM/c^2$ is the gravitational radius. Despite the use of selective refinement,
the three-dimensional nature of the problem,
along with the challenging simulation durations needed to adequately relax the disks, limited the number of frequency groups that could be
used to just twelve, spanning the six decades in energy from $5\times 10^{-3}$ to $5\times 10^3$ keV.

Of the original calculations (see Table \ref{tab:params}), one (S7Ea5) collapsed vertically
below our grid resolution due to the thermal instability.
Two others (S10Ea95 and S3Ea95) imposed inner grid boundaries outside
the ISCO, $r_{\text{min}} > r_{\text{ISCO}}$, potentially
affecting the development of the hot corona and by association the high energy spectral tail.
Hence we focus in this paper on the remaining four models: S01Ea0, S3Ea0, S3Ea75, and S3Ea9.
The first two intend to provide a comparison of different accretion rates with zero spin. 
The latter three compare the effects of different black hole spins at roughly the same accretion rate.

Table \ref{tab:params} summarizes the various simulations including their initial Novikov-Thorne target accretion rates in Eddington units ($\dot{m}_\mathrm{NT} = \dot{M}c^2/L_{\text{Edd}}$), 
spin in black hole mass units ($a_*$), final stopping times in gravitational time units ($t_G = GM/c^3$), and time-averaged mass accretion rate ($\langle \dot{m} \rangle_t$) measured over the interval $7500-10000 t_G$.
Note that most of the
runs develop quasi-steady mass accretion rates several times below their initial Novikov-Thorne target values \citep{Fragile23}. We also point out that the final stopping times are relatively short compared to global relaxation times. However, the computational demands
of these simulations prevented us from advancing them further. Regardless, and depending on model
and stopping times, the disks in most cases managed to achieve quasi-steady-state configurations out to about 40 or 50 $r_G$,
where maximum excursions of the infalling accretion rate have settled to well within a factor of two. This is one of the advantages of starting from the Novikov-Thorne solution, rather than an arbitrary torus.

\begin{deluxetable}{cccccl}
\tablecaption{Simulation Summary \label{tab:params}}
\tablecolumns{7}
\tablehead{
\colhead{Name} & \colhead{$\dot{m}_\mathrm{NT}$} & \colhead{$a_*$} &  \colhead{$t_\mathrm{stop}/t_G$} & \colhead{$\langle \dot{m} \rangle_t$} & \colhead{Notes}}
\startdata
S01Ea0  & 0.01 & 0    & 10,614 & 2.8 & $\cdots$ \\
S3Ea0   & 3    & 0    & 15,045 & 0.52 & $\cdots$ \\
S3Ea75  & 3    & 0.75 &  6,369 & $\cdots$ & $\cdots$ \\
S3Ea9   & 3    & 0.9  & 10,337 & 0.66 & $\cdots$ \\
S3Ea95  & 3    & 0.95 & 10,430 & 0.97 & $r_\mathrm{min} > r_\mathrm{ISCO}$ \\
S7Ea5   & 7    & 0.5  &  8,399 & 0.089 & Collapsed \\
S10Ea95 & 10   & 0.95 & 10,522 & 0.91 & $r_\mathrm{min} > r_\mathrm{ISCO}$ \\
\enddata
\end{deluxetable}

In this article we expand on our previous work by applying a new general relativistic Monte Carlo radiation transport code that we recently
developed \citep{Roth22} to post-process the four main cases at higher energy resolution and with an improved Compton scattering treatment
(our \mI results were computed in the Kompaneets approximation, where only the 
lowest order terms in $h \nu/m_e c^2$ are included in a Taylor expansion).
In addition to solving the full transfer equation (rather than a two-moment formulation of it), Monte Carlo (\mc) also naturally
allows for better estimates of escaping (outward directed) radiation compared to \mI, which instead incorporates
inward and outward directed contributions into a single net flux.

The rest of this paper focuses on describing the setup of our post-processed simulations (Section \ref{sec:methods}),
comparing \mI with \mc generated spectra while exploring effects of different accretion rates and
black hole spins (Section \ref{sec:disk_spectra}), generating radiation source maps of 
escaping \mc bundles (Section \ref{sec:radiationMaps}), and comparing simulated spectra together with observational data
(Section \ref{sec:comparison}).
Finally, we summarize our results in Section \ref{sec:conclusion}.
Throughout this work we have assumed a black hole of mass $M_\mathrm{BH} = 6.62 M_\odot$ with
a distance unit of $r_G = GM/c^2 = 9.8$ km and a time unit of $t_G = GM/c^3 = 3.3 \times 10^{-5}$ s.

\section{Methods}
\label{sec:methods}

Because our Monte Carlo treatment is currently limited to Cartesian meshes,
and our original RMHD simulations were performed on spherical-polar grids, we have to construct entirely new
grids for the post-processing calculations. 
High resolution is achieved near the black hole and along the disk midplane by using ratioed zoning that
spans more than four orders of magnitude in scale height. This approach increases cell sizes along
each axis by the geometric product $\Delta z_{i+1} = (1 + \epsilon) \Delta z_i$, where the constant ($\epsilon > 0$)
depends on the number of zones covering the length of the axis, recovered by iteratively solving
\begin{equation}
\label{eqn:ratiozoning}
\Delta z_{\text{min}} = \frac{\epsilon ~L_z}{(1+\epsilon)^{N_z} - 1} ~,
\end{equation}
for $\epsilon$ given a target resolution $\Delta z_{\text{min}}$ (smallest cell width), number of zones $N_z$, and grid length $L_z$.
The Cartesian grids cover the full spatial extent (with no reflection symmetries) and are designed to converge
onto the black hole center and midplane, with different geometric constants along each axis.
We chose mesh parameters to achieve resolutions comparable to the original \mI calculations,
with minimum cell sizes near the black hole of approximately
$(\Delta x, \Delta y, \Delta z)$ = $(0.06 r_G, 0.06 r_G, 0.0125 r_G)$,
compared to $(\Delta r, r\Delta \theta, r\Delta \phi)$ = $(0.05 r_G, 0.02 r_G, 0.05 r_G)$ 
at $r=2 r_G$ in the original simulations. 

\begin{figure}
\centering
\includegraphics[width=1.1\textwidth]{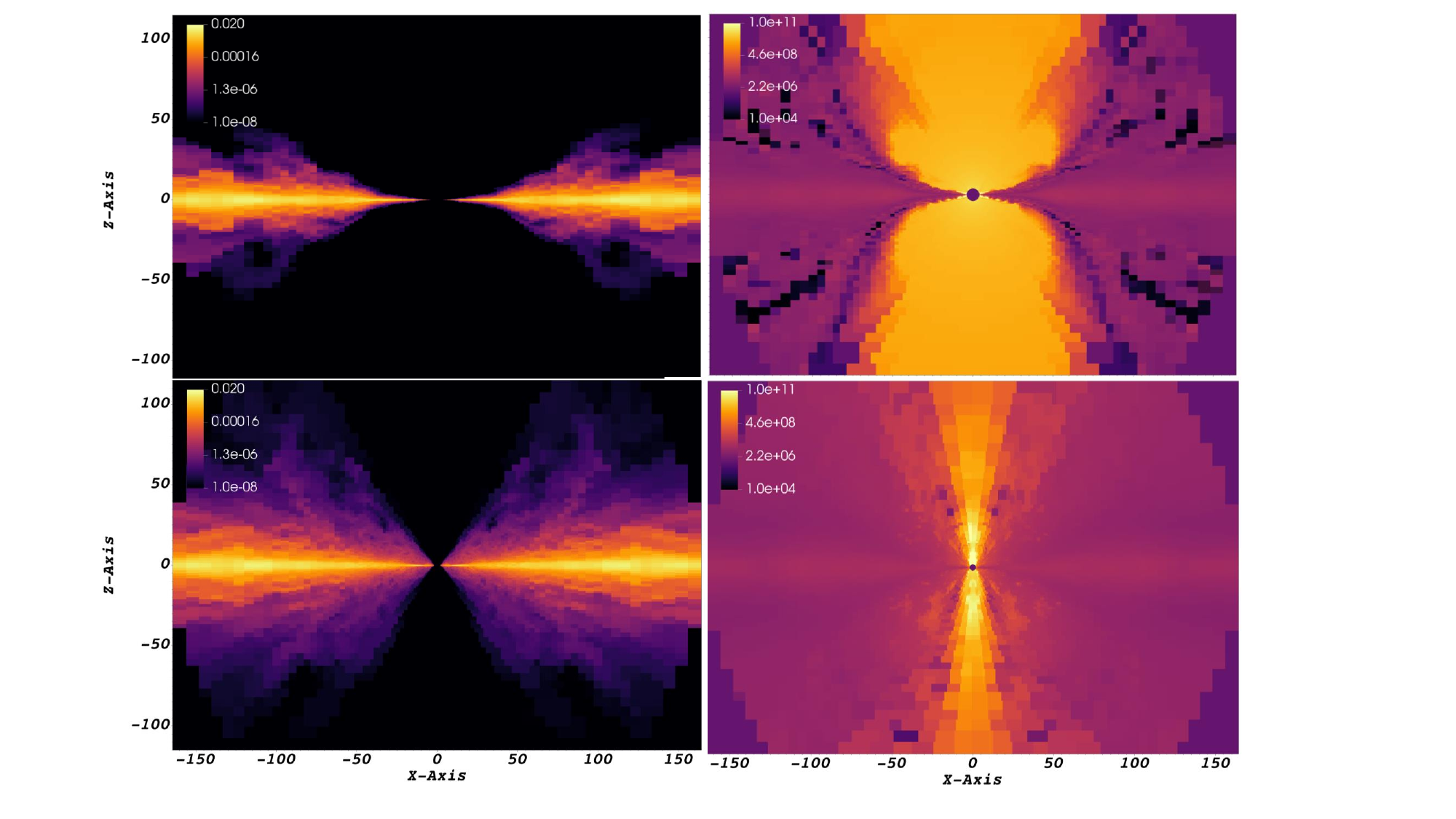}
\caption{
Logarithms of the gas density (left) and temperature (right) mapped onto a geometrically refined
Cartesian grid for models S3Ea0 (top) and S3Ea9 (bottom). 
The density is in units of g/cm$^3$,
temperature in Kelvin, and the data correspond to a single snapshot (not time averaged) taken at the last simulation times. 
The color scales are normalized to the same minima and maxima in both models.
The grid is designed to increase spatial resolution near the black hole (located at the origin) along the $x$ and $y$ axes, and
along the disk midplane along $z$, with minimum cell widths of ($0.06\times0.06\times0.0125) ~r_G$.
}
\label{fig:diskimages}
\end{figure}

All relevant geometric (spacetime metric) and hydrodynamic (density, temperature, velocity) quantities 
from the \mI simulations are mapped onto the new Cartesian grids
using a nearest-neighbor, first order procedure, then rotated by azimuthal symmetry to fill
the volume (the original curvilinear calculations were performed with $\pi/2$ periodicity in $\phi$).
Figure \ref{fig:diskimages} shows the mapped density (left) and temperature (right) for models S3Ea0 (top) and S3Ea9 (bottom),
representing the two most different (quasi-steady state) disk configurations. 
The S3Ea9 disk with a spinning black hole has a significantly thicker
vertical extent (evident in the density map) up close to the black hole, and a substantially hotter (but also narrower) coronal region reaching
well beyond the disk scale height. These are both expected, as spinning black holes have been shown to drive much stronger outflows than their non-spinning counterparts \citep{Beckwith08, Tchekhovskoy10}.

The hydrodynamic data are used to source photon bundles that are subsequently geodesically transported in 3D using the 
super-photon Monte Carlo method described in \citet{Roth22}. Since we treat these photon bundles as
diagnostic particles, not solving the fully coupled radiation-hydrodynamics equations, we ignore Fleck
scattering in the calculations but add geodesic curvature, gravitational redshifts, 
free-free emission and absorption, and Compton scattering of photons by electrons.
Photon mean-free-paths and interactions are computed through the use of locally orthonormal tetrads
constructed in the fluid rest frame, accounting for Doppler shifts (measured in the
coordinate frame) due to fluid motions. The accuracy of Compton scattering is improved by incorporating
high temperature corrections to the fluid frame mean-free-paths (via table lookups). 
For scattering events, we sample electron velocities from a relativistically corrected thermal distribution (Maxwell-Juttner),
modeling effects of the relative alignment of the electron and photon momenta, as well as the photon energy 
measured in the electron rest frame, on the scattering rate. 
We boost into the sampled electron rest frame to perform the scattering, with an outgoing photon direction 
and energy shift sampled according to the Klein-Nishina differential scattering formula, 
before finally transforming back to the coordinate frame to continue propagating the photons. 
For further details, consult section 3.6 of \citet{Roth22}. 
We note that our current Monte Carlo implementation does not account for stimulated Compton nor
double Compton emissions, which may be relevant at very high temperatures and low plasma densities \citep{Thorne81,McKinney17}.

As bundles propagate, they have a chance of being absorbed by electrons via the free-free process, rather than scattered, 
with a probability that relates to the ratio of the absorption and scattering opacities. 
In the case of absorption we stop tracking the bundle. For surviving bundles, we do not modify the bundle weight 
(the number of photons represented by the bundle) due to absorption along the bundle's path. 
Therefore, we rely on sourcing a large number of bundles such that enough of them can escape to produce adequate 
statistics in each range of frequencies in the final spectrum. We follow this approach because it more closely tracks the 
particular method developed in \citet{Roth22} for use in coupled radiation-hydrodynamics calculations. However, it can lead to 
noisy regions of the computed spectrum, as will be discussed below. In the future we may wish to explore the option of 
deterministically attenuating the bundle weight to account for absorption processes, rather than this stochastic, all-or-nothing 
approach. This would increase the compute time per bundle tracked, 
but may reduce the noise in certain regions of the spectrum.

Fixing the grid resolution ($\Delta z_{\text{min}}$ in equation (\ref{eqn:ratiozoning})) 
and incorporating the physics interactions described above to match the \mI calculations, 
it remains to explore the effects of the following parameters in the post-processing analysis:
(1) the number of zones covering each axis, 
(2) the number of photon bundles,
(3) the number (and choice of) sequential time dumps from the \mI calculations to post-process, 
(4) the physical time over which the \mc bundles are tracked,
(5) the maximum number of single interaction steps allowed before we stop tracking a bundle, and
(6) the minimum radius a photon can reach before we stop tracking it.

The motivation for the first two parameters derives from the fact that it is not enough to 
simply define the smallest cell size near the black hole or along the disk midplane.
It is equally important to resolve the hot, relatively lower density material above the disk scale height
which can affect the spectrum at energies just above the soft thermal peak. This is an especially difficult portion of the spectrum to
resolve properly as it depends on the grid resolution some distance off the midplane and adequate
photon statistics in cells emitting photons primarily at energies in the exponentially diminishing thermal tail of the spectrum.
Both are addressed by increasing the number of zones together with the number of photon bundles in these regions, 
which in three dimensions is particularly challenging, even for post-processing. After performing numerous convergence studies we
eventually settled on $192^3$ cells, 5 total bundles emitted per zone each Monte Carlo cycle, and 20 \mc cycles between
data dump intervals for a total of $\sim2\times10^{9}$ total bundles emitted over a typical calculation
across the last three dump intervals.
In general, temporal and statistical count variances depend strongly on photon frequency, so we plot
time variances together with each spectrum as they are presented in Sections \ref{subsec:disk_nospin} and 
\ref{subsec:disk_spin}, and discuss statistical count variances separately in Section \ref{subsec:variance}.

The motivation for the next couple parameters comes from our observations that the \mI radiation-hydrodynamics calculations have not fully relaxed 
to steady-state, even after $10^4$ dynamical times. As mentioned above, and for the purpose of evaluating transient effects on
emission features, we settled on postprocessing the last three data dumps from each model, separated 
by an interval equal to the ISCO orbital period, $t_{\text{ISCO}}\sim92~t_G$. 
These time intervals are additionally sub-divided into 20 sub-cycles that are used for sourcing photons in the intervening time. 
This process of reading, mapping, sourcing, and evolving is repeated until the last of the three cycles completes.  
We compared this to an alternative approach of post-processing the final (single) radiation-hydrodynamics data dump until 
the radiation field reaches a steady-state at roughly the 1\% level in total luminosity, verifying that while this leads to a 
small increase in the overall luminosity (typically a few percent) 
it does not alter any of our conclusions regarding the shape of the spectrum. Of course, we cannot guarantee that the 
spectrum does not evolve on a longer time scale without actually evolving the radiation-hydrodynamics calculations out to longer times.

Choices for the final two parameters are motivated by practical limitations on computing time. Photons trapped in optically thick regions
can significantly slow the calculations if allowed to evolve unconstrained. 
We therefore impose a condition to drop photons if they exceed $5\times10^4$ 
interactions per cycle. We also drop photons if they approach within a cell width of the event horizon,
where the tetrad (and inverse) operations become sensitive to strong curvature, increasing errors
in the transformed 4-momenta normalizations. Dropping these rare trajectory photons does not visibly affect the spectra.

An important distinction between the \mI and \mc spectra is the latter's ability to distinguish between incoming and outgoing radiation. 
Radiation fluxes in the \mI radiation-hydrodynamics simulations can be highly dynamic, occasionally even switching directions (toward or away from the black hole),
resulting in alternating periods of positive and negative net radial fluxes, especially during the early relaxation phase. 
Because the \mI closure scheme only tracks the net flux
in any given energy bin and grid cell, it cannot distinguish incoming from outgoing radiation, and cannot be entirely trusted to represent
true output flows except perhaps once steady-state conditions are reached. As we noted earlier, none of our calculations were run long enough
to have completely settled the disk; hence, we anticipate (and observe) some differences in
\mI and \mc spectra that we attribute to relaxation and transient corona heating effects.

\subsection{Spectra}
\label{sec:SpectralBinning}

The spectra that would be observed from our model systems are dependent on time, viewing angle, 
and approximations introduced by our numerical methods as well as physics inputs.  This includes the
spectral binning structure used to collapse and accumulate energy tallies or fluxes, a five-dimensional
object composed of time, polar angle, azimuthal angle, photon energy (frequency), and radius.
Due to this large number of dimensions, we settled on
15 bins to represent time spanning the duration of each calculation (thus, five bins per dump analyzed),
10 uniformly spaced (in $\cos\theta$) polar bins covering $0 \le \theta \le \pi$,
and 1 azimuthal angle bin averaging over all angles.
For \mI, the number of energy bins is set to 12, logarithmically spaced across 5 - $5\times10^6$ eV corresponding to the same bin number
and range as the original radiation-hydrodynamics calculations. For \mc, the number of bins is increased to 50 covering 
a greater range 1 - $10^8$ eV.
We use the same number of radial bins (10) for \mI as \mc, but because this data is used differently in the two methods (as we discuss below)
the ranges differ: $20 r_G$ to the outer boundary for \mI, and from the event horizon to the outer boundary for \mc.

\subsubsection{\mI spectral binning}

The spectral extraction procedures differ between \mI and \mc
due to their fundamentally different treatments and representations of the radiation field.
\mI spectra are reconstructed from evolved fluxes over (radial) surfaces parameterized by the five-dimensional grid. 
In order to extract spherically directed fluxes from the Cartesian mesh, we construct a spherical mesh composed of
$50\times50$ ($N_\theta \times N_\phi$) angular cells at each radial bin location, overlaid onto the Cartesian mesh.
The spectral luminosity is projected from the Cartesian (evolution) mesh onto the spherical (spectral) grid at each time and radius,
then integrated across user-specified angular domains ($\delta\theta, \delta\phi$) as follows
\begin{eqnarray}
L_\nu (t, r, \delta\theta, \delta\phi) &&      = - \int_{\delta\phi} \int_{\delta\theta} \sqrt{-g} R^i_{t(\nu)} dA_i \Delta\nu~, \\
      &&\approx -\frac43 \sum_{\delta\phi} \sum_{\delta\theta} \sqrt{-g} u_R^0 u_{0 R} E_R v^r_R \Delta\theta \Delta\phi \Delta\nu ~,
\label{eq:luminosity}
\end{eqnarray}
where $dA_i$ ($\Delta\theta ~\Delta\phi$) represent the orthogonal (to the radiation flux) cell face areas in curvilinear coordinates,
$R^i_{t(\nu)} \equiv R^r_{t(\nu)}$ is the radiation flux in a frequency bin of width $\Delta \nu$ centered at frequency $\nu$ with a radial
component approximated by the radiation primitive energy $E_R$ and radial velocity $v^r_R$ as
$R^r_{t(\nu)} \approx (4/3) u_R^0 u_{0 R} E_R v^r_R$ \citep{Anninos20}.

One can either use the spectral angular grid to define the direction of the radiation outflow, or the primitive radiation velocity.
The former is a reasonable approximation for mostly spherical flows. The latter is more appropriate from a viewing angle perspective,
but is itself an approximation if the extraction domain is optically thick (or strongly gravitating). However because we
extract spectra through relatively narrow cones centered along the optically thin funnel regions (directed along the poles),
this is not a major concern. In either case we have verified that both options yield essentially
identical results.

\subsubsection{\mc spectral binning}

Unlike \mI, Monte Carlo spectra are not reconstructed from flux surface integrals. Photon bundles are instead only
tallied and their attributes projected on the 5D spectral grid 
when they cross the outer boundaries of the computational domain. The direction  angles in this case correspond to
bundle propagation directions when they exit the grid, regardless of the spatial position at which they escape.

Radial bins for \mc have a different meaning than for \mI. Here they represent the locations where bundles are first birthed, not
radial positions of shells where fluxes are measured. Birth zone binning allows for the flexibility of eliminating
certain regions in the spectral energy collection process, and thereby effectively ignoring contributions from bundles birthed
in those regions. This technique is especially useful for determining the source and effect of regional emissions on the spectrum 
(as we discuss in Section \ref{sec:radiationMaps}). Because we are particularly interested in defining the source of high energy
photons, we extend the radial bins to cover the entire computational domain from the event horizon to outer boundary (unlike \mI where
the radial bins started at $20 r_G$, roughly where the disks have achieved a quasi-steady-state).

\subsection{Kompaneets Approximation}
\label{subsec:kompaneets}

The \mI method deals with Compton scattering by solving the Kompaneets equation
approximating binary collisions as a second order differential operator in the isotropic and low energy (non-relativistic) regimes \citep{Kompaneets57,Thorne81}:
\begin{equation}
\frac{\partial E_\nu}{\partial t} = \sigma_0 ~x \frac{\partial}{\partial x}
                                    \left[ g_\nu(\nu,T_e) x^4 \left(  \theta \frac{\partial}{\partial x} \left(\frac{E_\nu}{x^3}\right)
                                                           + \frac{E_\nu}{x^3} \left(1 + \gamma\frac{E_\nu}{x^3}\right) 
                                                     \right)
                                    \right] ~,
\end{equation}
where $E_\nu$ is the radiation energy, $\sigma_0=N_e \sigma_T c$, $N_e$ is the electron number density, 
$\sigma_T$ is the Thomson scattering cross section,
$g_\nu(\nu,T_e)$ are relativistic (frequency- and temperature-dependent) corrections \citep{Cooper71}, $x=h\nu/m_e c^2$,
$\theta= k_b T/m_e c^2$, and $\gamma=(hc/m_e c^2)^3/8\pi$. By contrast our \mc method makes no such simplifications in either energy
or distribution functions, and is equally valid in relativistic (hot) environments. 

It is prudent therefore to compare and assess the accuracy
of the Kompaneets approximation in the hot environments experienced by photons in our accretion disk models.
For this purpose we construct a simplified spherical model composed of concentric layers of material at analogous densities and temperatures.
In the core we represent the dense disk as a sphere with density $10^{-3} ~\text{g/cm}^3$, temperature $5\times10^6$ K, and radius $10 r_G$.
The core is surrounded by a hot corona at lower density $10^{-5} ~\text{g/cm}^3$, hotter temperatures  $10^8$ - $5\times10^8$ K, and
outer radius $20 r_G$. Outside the corona we have background gas at even lower density $10^{-8} ~\text{g/cm}^3$ in thermal equilibrium with the corona.
We evolve this test model as a pure radiation transport problem, with no coupling to the hydrodynamics (other
than the temperature source). Our purpose is to
compare the output spectra computed by the two transport models, \mI and \mc, under conditions that give rise to an upscattered energy
distribution (via inverse Compton scattering through the corona) matching our disk models.
The models are evolved on a $64^3$ grid and run long enough for the total (frequency-integrated) luminosity to converge to within 1\%.

\begin{figure}
\centering
\includegraphics[width=0.6\textwidth]{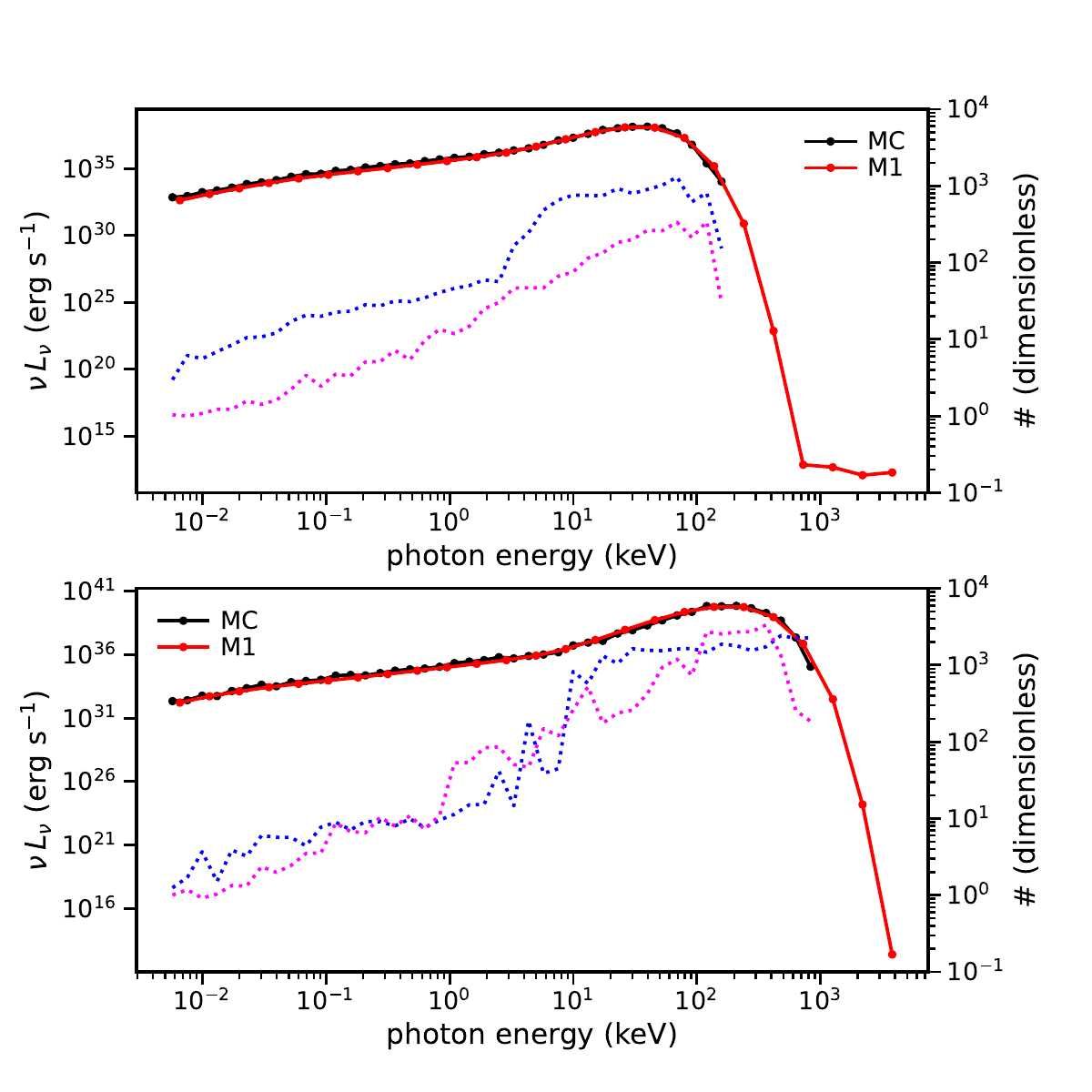}
\caption{
Spectral luminosities from two spherical models testing the accuracy of the Kompaneets approximation 
in resolving inverse Comptonization through hot coronal regions.
The top (bottom) plot shows the spectrum with a corona temperature of $10^8$ K ($5\times10^8$ K), producing Compton
peaks corresponding roughly to what we observe in our disk models.
In both images, the Kompaneets solution is shown in red and the full \mc transport in black.
Against the right axis, we also plot the energy-averaged number of scatters (blue dotted line) and
energy increase factor (magenta) experienced by the \mc bundles in each bin.
}
\label{fig:kompaneets}
\end{figure}

The results are plotted in Figure \ref{fig:kompaneets} where we compare spectra from \mc (in black) against \mI (in red)
representing the radiation flux exiting the outer grid boundaries summed over all angles.
We also plot in blue the number of scatters the packets undergo, using an average weighted by the packet energy when the packet escapes, and referenced to the (dimensionless) axis label on the right. In magenta, we plot the factor by which packets escaping in each frequency bin have had their fluid-frame energy changed by Compton scattering effects. The magenta curve also corresponds to the axis label on the right. 
We show two cases differing only by the corona temperature: the top image corresponding to a 
corona temperature of $10^8$ K, the bottom to $5\times10^8$ K.
Notice the inverse Compton peaks, at roughly 50 and 200 keV respectively, are comparable to the Compton peaks observed in
our disk models (see figures in Sections \ref{subsec:disk_nospin} and \ref{subsec:disk_spin}). 
Agreement in frequency-integrated luminosities between \mI and \mc is excellent, better than a few percent and comparable to the convergence
criteria used to terminate the calculations.

\section{Disk spectra}
\label{sec:disk_spectra}

\subsection{Zero Spin}
\label{subsec:disk_nospin}

\begin{figure}
  %
  \gridline{
    \leftfig{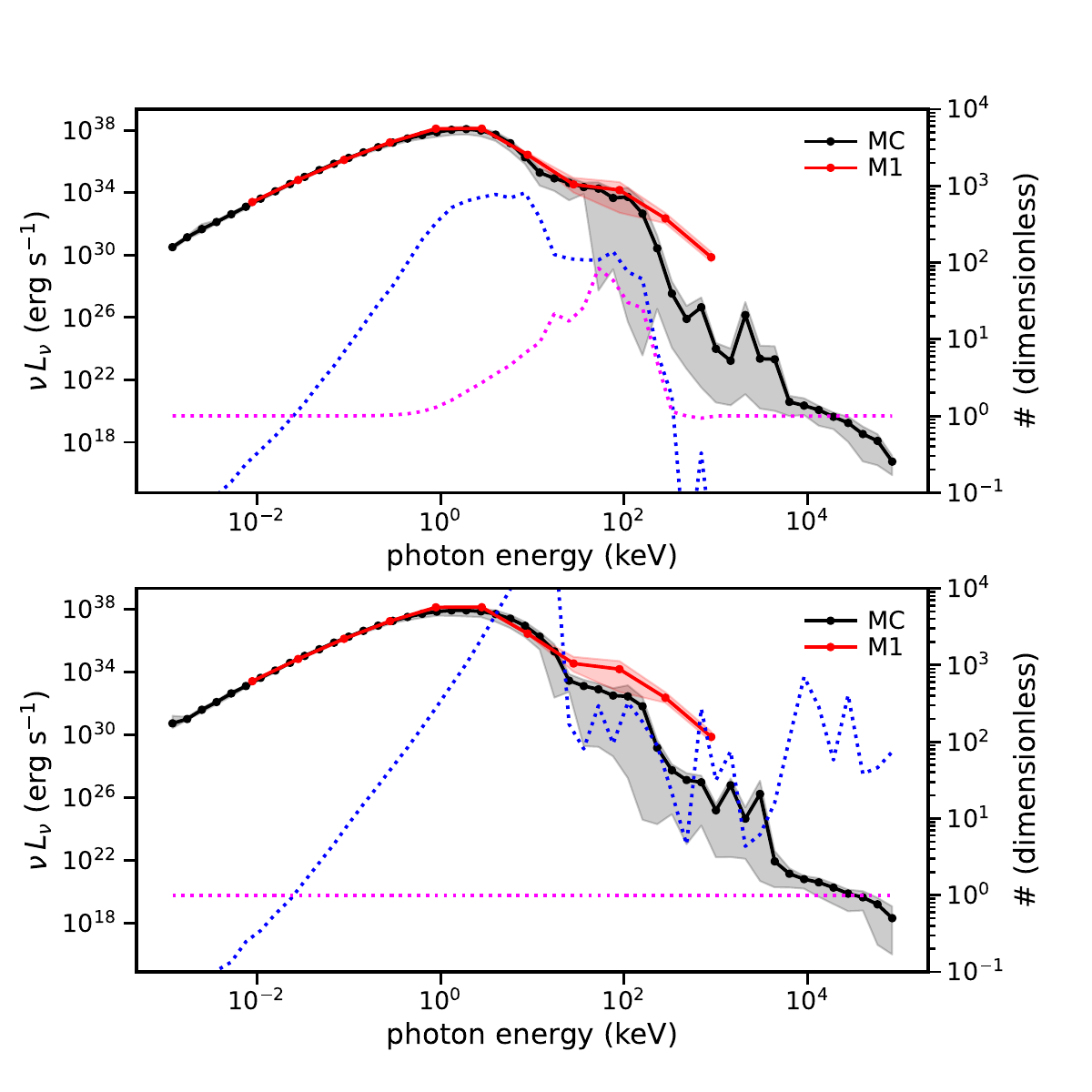}{0.5\textwidth}{(a)}
    \rightfig{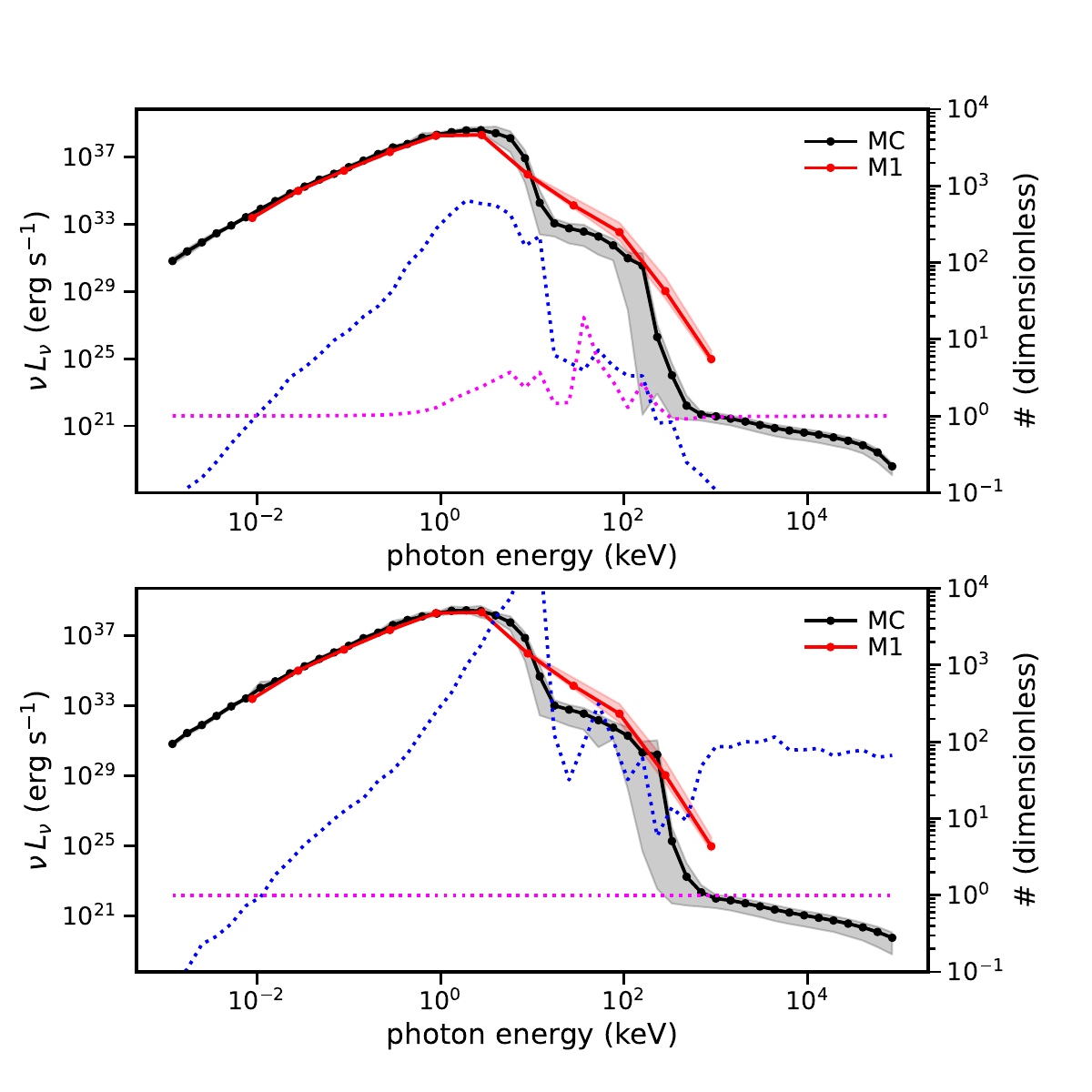}{0.5\textwidth}{(b)}
    }
  \caption{
    Spectral outputs derived from inlined \mI (red) and post-processed \mc (black) calculations. Column (a) corresponds to the S01Ea0 disk model, while column (b) corresponds to the S3Ea0 model. The top plot of each column includes Compton scattering, while the bottom plot replaces Compton scattering with Thomson. In all panels, the spectra represent the effective $4\pi$ outputs estimated from the flux traveling
through a $54^\circ$ half-angle cone aligned with the north pole.
    Shaded regions represent time variances in the \mc results across the last 5 time bins, cumulatively corresponding to roughly one ISCO orbital period.
    As in Figure \ref{fig:kompaneets}, the blue and magenta dotted lines represent the 
energy-averaged number of scatters and energy increase factor
experienced by the \mc bundles in each energy bin relative to the dimensionless scale on the right axis label.
}
\label{fig:spec_cases1and2}
\end{figure}

We begin discussions of the disk simulations with the two non-spinning cases: S01Ea0 and S3Ea0.
Figure \ref{fig:spec_cases1and2} compares the \mI (red) and \mc (black) effective $4\pi$ 
spectral outputs from S01Ea0 and S3Ea0 in columns (a) and (b), respectively. Also plotted are the energy-averaged number of scatters (in blue) and the 
energy increase factor from inverse Compton scattering (in magenta) experienced by the \mc bundles.
The two separate panels in each column correspond to different scattering treatments: Compton (top) and Thomson (bottom).
For \mI, we collapse (and average) fluxes from the five outermost radial bins at radii $>100 r_G$.
The spectra are constructed by evolving emissions through the three consecutive time dumps ending at $\sim10^4 t_G$, 
and recording the spectra from the first four polar angle bins corresponding to a 
viewing angle of $54^\circ$ half-angle cone oriented along the pole axis  
and projecting it to the full $4\pi$ solid angle.
In other words, these luminosities correspond to what an observer would infer as the isotropic-equivalent luminosity
for the accretion disk system if viewed normal to the disk.

The shaded regions in Figure \ref{fig:spec_cases1and2} represent temporal variances (or more precisely here, excursions) across the last five time bins corresponding to a total of about one ISCO period 
(as a reminder, we use 15 time bins to cover 3 simulation dump cycles, each cycle separated by 1 ISCO period).
Notice the low-energy thermal portions of the spectra are essentially relaxed, exhibiting little variance. However
the high energy parts remain highly dynamical and unsettled, in both the \mc and \mI simulations. 
In addition to limited bundle statistics, this might reflect short time-scale variability that persists over the evolution of the disk. 
We point out that bundle statistics are poorest at photon energies above 100 keV because 
only a relatively small fraction of bundles are scattered to these energies, and only a small fraction of those escape without being re-abosrbed.
Shot noise can be reduced to some extent by increasing the spatial resolution (number of cells covering the disks and coronae)
and increasing the number of photon bundles, which we have done for several cases to arrive at our current
parameter choices, compromising somewhat on variance for computational expediency.

Despite the globally unsettled nature of the simulated environments, it is comforting to see
good agreement between the \mI and \mc results for both cases that include Compton upscattering. They match across the settled thermal spectrum
nearly perfectly, and, considering the coarsely binned energy grid used in the \mI calculations, 
both methods exhibit similar (weaker) fluxes at energies between 10 and 100 keV.
The total, frequency-integrated luminosity amplitudes agree to roughly 35\% at the time exhibited:
$0.25 L_\mathrm{Edd}$ and $0.88 L_\mathrm{Edd}$ for the S01Ea0 and S3Ea0 models calculated by \mc, respectively.
There is some disagreement in the shape of the spectrum in S3Ea0 (column (b) of Figure~\ref{fig:spec_cases1and2}) 
at photon energies between approximately 30 and 100 keV, where the \mI flux exceeds that of the \mc.  
While we found excellent agreement with our test problem in Section \ref{subsec:kompaneets} between the Kompaneets 
approximation and the full \mc solution, we did not explore all relevant density and temperature regimes, 
or the presence of strong density and temperature gradients, with such an idealized problem. 
These differences might also be related to differences in spectral fidelity or binning procedures 
between the two methods (Section~\ref{sec:SpectralBinning}).

Although S01Ea0 and S3Ea0 both produce dominant thermal peaks between 2 and $2.5$ keV, and hotter but weaker emissions up to $\sim100$ keV,
case S3Ea0 clearly exhibits much lower Compton energy exchanges as evidenced by the lower flux levels, and by
comparing the dotted lines in the top panels of Figure \ref{fig:spec_cases1and2}.
It was our hope to have used these two cases to study the
effects of varying $\dot{m}_\mathrm{NT}$ on spectral outputs, without the complication of spin.
However, and as we noted in \citet{Fragile23}, case 
S3Ea0 was affected by the thermal instability significantly more than the other cases considered here, with the inner regions ($<20 r_G$)
collapsing by roughly a factor of two from its initial configuration measured by the disk scale height
\begin{equation}
\langle H(R)\rangle_\rho = \sqrt{\frac{\int \rho^2 z^2 \mathrm{d}V}{\int \rho^2 \mathrm{d}V}}~.
\label{eqn:height}
\end{equation}
Spectral differences at these intermediate frequencies might instead be attributed to differences in the relaxation behavior of the 
inner regions of the disks, driven by the thermal instability more so than $\dot{m}$. 

For case S3Ea0, the 10-100 keV X-rays come from direct emissions associated with the collapsed inner regions,
which produce a slightly cooler disk due to the relatively enhanced rate of cooling experienced by this model.
More significantly, S3Ea0 maintains a substantially cooler corona, by roughly an order of
magnitude in regions near the black hole and immediately beyond the disk scale height,
severely limiting Compton interactions. This is confirmed
by the low Compton exchange factor (the magenta curve) in the top plot of column (b) of Figure \ref{fig:spec_cases1and2}, and by the similarity of that plot with the bottom plot displaying the spectrum when Compton scattering is replaced by elastic scattering in the postprocessed simulations.
The top and bottom spectra are essentially identical, differing only in the number of scatters, which is expected
when photons are prevented from upscattering, and the Compton energy factor, which for elastic scattering is unity across the energy grid.
Contrast that with the corresponding panels for case S01Ea0 in column (a), 
where the inclusion of Compton upscattering in the
\mc calculations (top panel) has a significant and observable effect between 10 and 100 keV, matching the \mI
output nicely despite the Kompaneets approximation.

\subsection{Nonzero Spin}
\label{subsec:disk_spin}

\begin{figure}
\centering
%
  \gridline{
    \leftfig{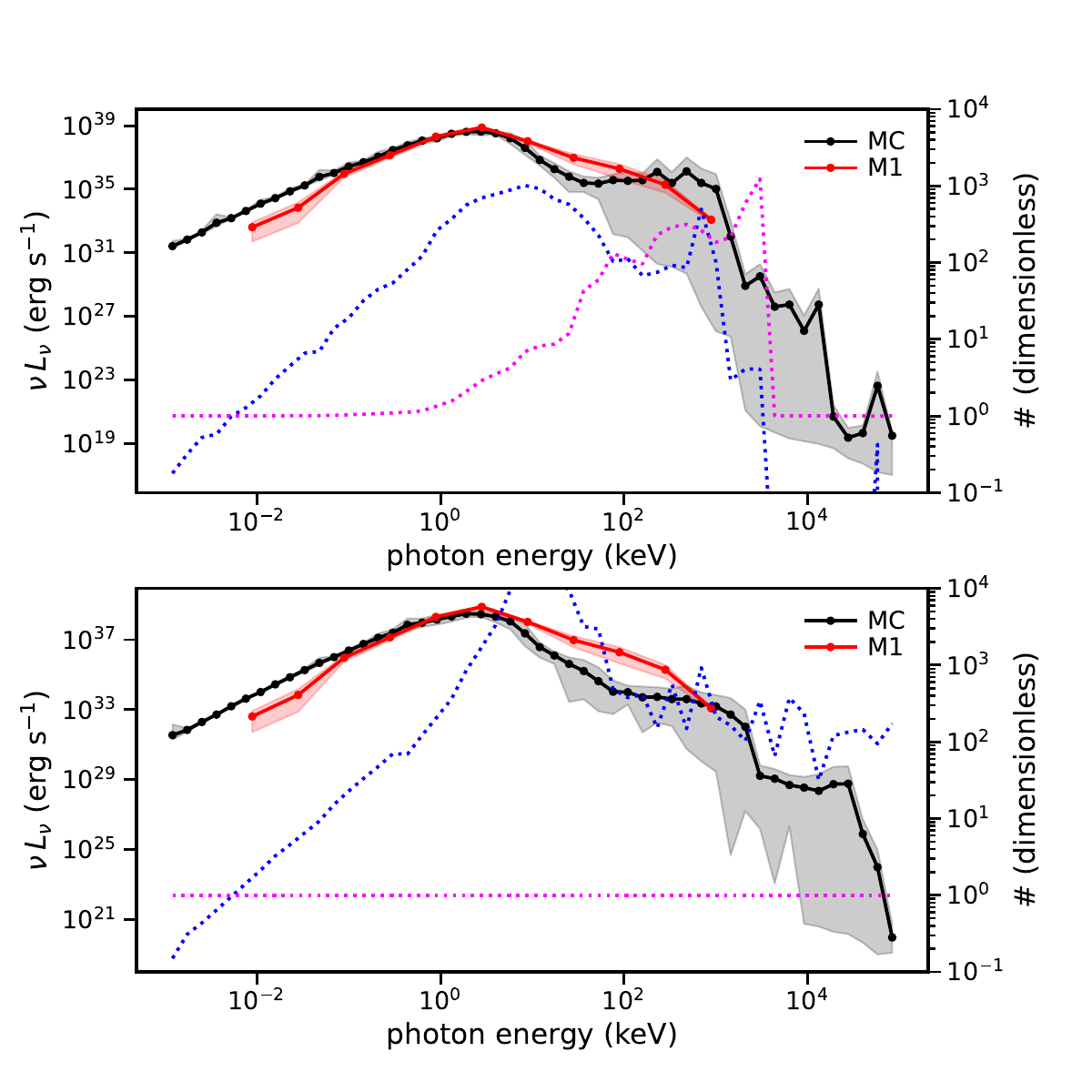}{0.5\textwidth}{(a)}
,    \rightfig{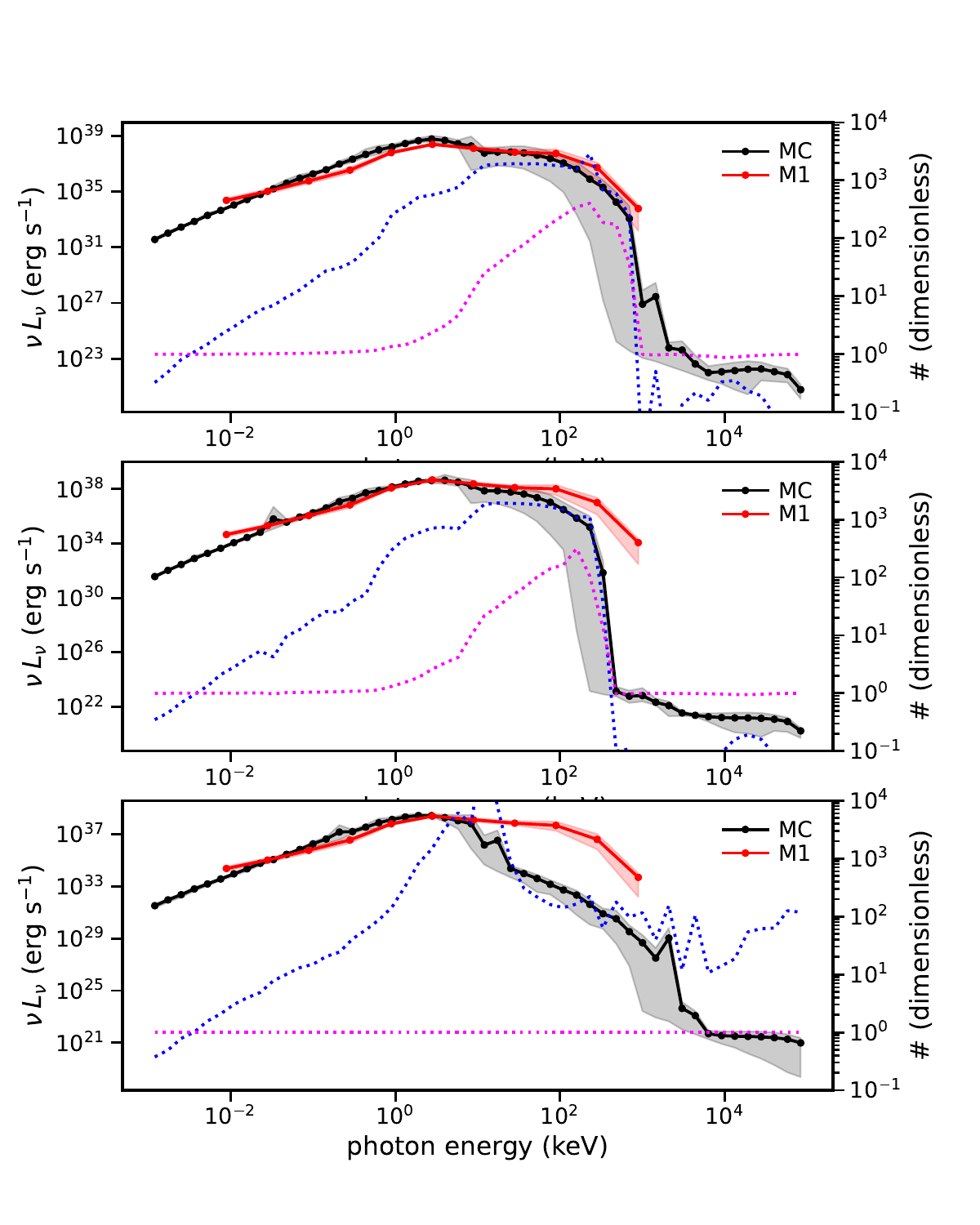}{0.5\textwidth}{(b)}
    }
\caption{
Similar to Figure \ref{fig:spec_cases1and2}, where now column (a) corresponds to the S3Ea75 model and column (b) corresponds to S3Ea9. For S3Ea9, the extra middle panel shows the Monte Carlo spectrum when we ignore emission from the potentially problematic, and numerically adjusted for stability, regions of the spin-aligned funnel.
}
\label{fig:spec_cases7and6}
\end{figure}

We considered a range of black hole spins from $a_* = 0$ to 0.95 (see Table \ref{tab:params}) but limit our discussions
here to the three cases not affected by computational deficiencies: S3Ea0 ($a_*=0$), S3Ea75 ($a_*=0.75$), and S3Ea9 ($a_*=0.9$).
In general, we expect the higher spin cases to exhibit increasingly harder spectra for the same $\dot{m}$, 
simply because the inner radius of the disk will move closer to the black hole, leading to a higher maximum temperature. 
This has been demonstrated in previous work by post-processing MHD simulations \citep{Kinch21} and is also what we find when we
compare spectra from our multi-group RMHD models in column (b) of Figure \ref{fig:spec_cases1and2}, column (a) of Figure \ref{fig:spec_cases7and6}, and column (b) of Figure \ref{fig:spec_cases7and6}, corresponding to spins 0, 0.75 and 0.9, respectively.
The time and region bins used for these figures are the same as in the previous (zero spin) plots, but because we observe significant
asymmetries above and below the disk, we average the outgoing spectra in the spin cases through both the positive
and negative pole axes (the $54^\circ$ conical half-angle remains the same). 

As for the nonspinning cases, 
we present spectral outputs for both Compton (top panels) and Thomson scattering (bottom panels) in Figure
\ref{fig:spec_cases7and6}, for case S3Ea75 in column (a) and S3Ea9 in column (b).
We demonstrated in the previous section that upscattering in the $a_* = 0$ models can dominate emissions at frequencies between
10 and 100 keV; the addition of spin exaggerates that dominance and extends it to even higher energies.
In fact, most of the X-ray emissions from about 30 to several hundred keV are attributed to inverse Compton scattering.
Notice how the spectra get progressively harder (the peak moves up and to the right) as the spin increases. 
However, the prominent thermal peaks do not vary much, increasing only slightly as a function of spin (see also Figure~\ref{fig:spectra_all}).

We had speculated in \citet{Fragile23} that radiation in the high energy bands might have been artificially high due to the 
approximate (Kompaneets) treatment of Compton scattering in our \mI scheme.
Instead, we observe reasonably good agreement with \mI in terms of the total luminosity emitted at photon energies 
greater than 10 keV, even though the exact shape of the spectrum differs at those photon energies, particularly for the S3Ea75 calculation. 

We also observe a small
but nonetheless significant effect from our treatment of the low density funnel in the original \mI
calculations where we applied energy corrections  in regions where the ratio of magnetic to thermal pressure becomes
high enough to trigger numerical instabilities. These corrections can result in artificially high temperatures
potentially hardening spectral outputs. This is evident comparing the top and middle panels in column (b) of Figure \ref{fig:spec_cases7and6}:
the top panel is generated by including the contribution of all cells to the spectrum; the middle panel ignores
emission from regions where the magnetic pressure exceeds 24 times the gas pressure, the condition for triggering the corrections;
finally, the bottom panel, as in all the other spectra, is the result we get when we replace Compton scattering with Thomson.

Our calculations (spinning and nonspinning) indicate Comptonization to be less 
relevant for the disk regions sourcing photons at energies greater than several hundred keV. 
Above those energies we find thermal free-free emissions emanating from the very hot, low density, optically thin gas 
found in the funnel regions close to the black hole. Those conditions
provide the appropriate emissivity to achieve such high energies given the physical processes we are modeling. 
Photons sourced from these hot coronal regions propagate essentially
freely to the outer boundaries leaving smooth (unaffected by scattering) spectral signatures above $10^4$ keV.
This is supported by the dotted blue curve in the top panel of column (b) of Figure \ref{fig:spec_cases7and6}, which shows photons at these
high energies escape with minimal scattering. On the other hand, the blue and magenta curves together 
indicate that low-energy photon bundles scatter with little to no effect on their energies,
suggesting that photons contributing to the soft thermal component scatter with high probability in
relatively low temperature gas.
As noted in Section~\ref{sec:methods}, our statements concerning the highest photon energies
are subject to the caveat that we have neglected double Compton interactions, which can dominate
free-free emission from very hot disk regions \citep{McKinney17}.

Notice the high energy part of the spectrum from case S3Ea75 appears different from the other models. It suffers a higher degree
of Monte Carlo noise, swamping even coronal free-free emissions at $\sim10^4$ keV. The \mI result also suffers greater
variance in the thermal portion of the spectrum, and a significant drop in luminosity at $10^{-2}$ keV.
This is a strong indicator that this model is further from steady-state than the other calculations, or that emission 
from this portion of the disk exhibits variability on time scales shorter than those tracked by post-processing.
Indeed, as Table \ref{tab:params} shows, S3Ea75 was run to just slightly more than half the physical time of the other calculations 
and remains in a highly dynamic transient state at $t \sim 6\times10^3 t_G$. The drop at $10^{-2}$ keV, for example, is attributed to
the radiation flux switching directions relative to the black hole at certain locations and times.
This is typical of the radiation flow in many of our calculations at early times, particularly in the low energy bins.
Although S3Ea75 is not representative of steady-state emissions, it remains a useful benchmark comparing transport methods, 
in addition to demonstrating the gradual hardening of the spectrum with increasing spin.

We additionally point out that for a given mass accretion rate, higher spins should lead to higher luminosities. 
Again, this is because as spin increases, the inner disk radius moves closer to the black hole, allowing more gravitational binding energy 
to be released before matter plunges into the black hole. Our results are consistent with this expected behavior, finding
(frequency-integrated \mc) effective isotropic luminosities of 0.9, 1.0, and 1.3 $L_{\text{Edd}}$ from S3Ea0, S3Ea75, and S3Ea9, respectively.

\subsection{Statistical Variance}
\label{subsec:variance}

\begin{figure}
\centering
\includegraphics[width=0.6\textwidth]{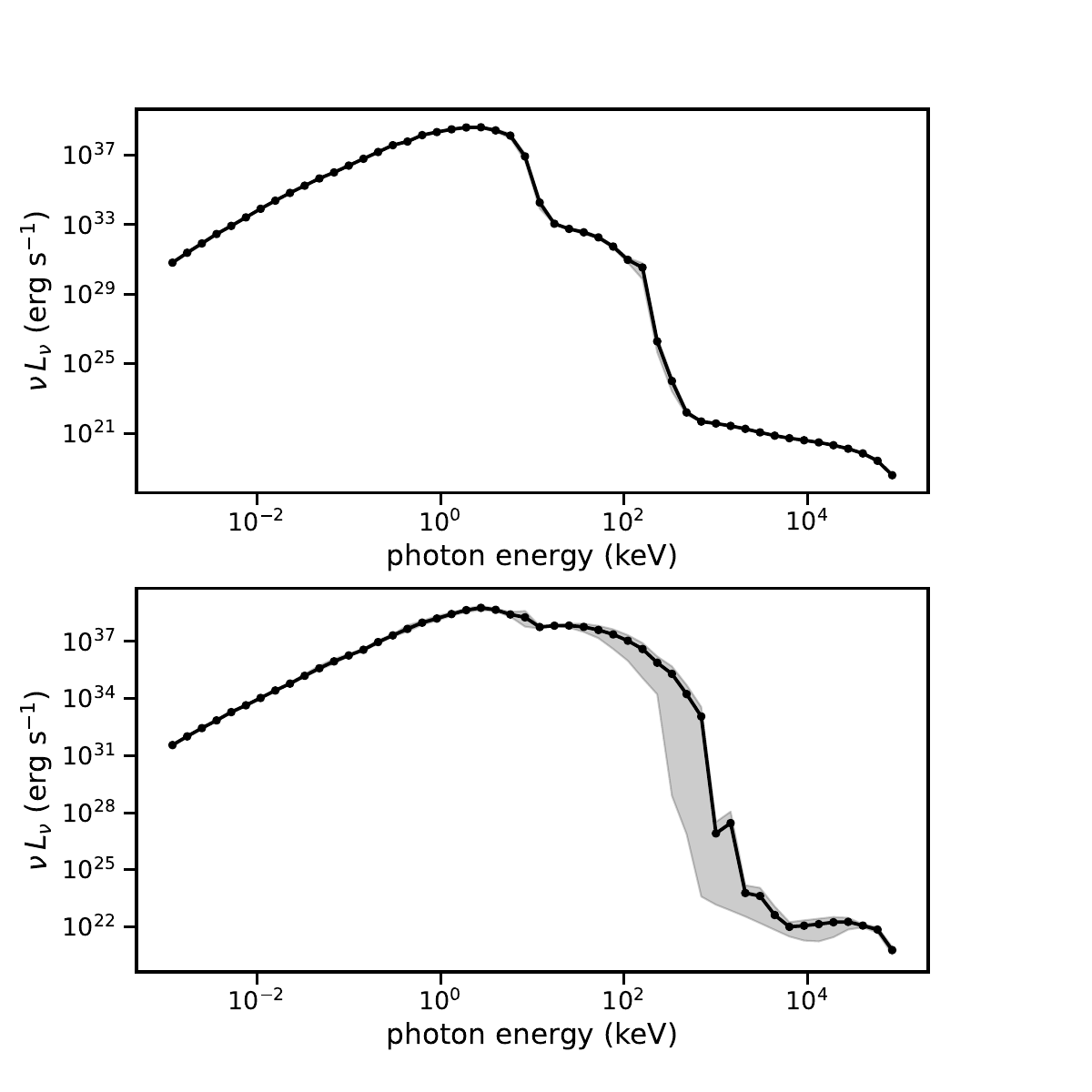}
\caption{
Statistical variance plots for cases S3Ea0 (top) and S3Ea9 (bottom), showing the Monte Carlo spectra along with 1$\sigma$ contours
representing accumulated shot noise. The spectra in both cases are computed from a total of more than $10^7$ bundles
across the entire energy grid.
}
\label{fig:spec_variance}
\end{figure}

All spectra presented thus far have superimposed the range of fluctuations observed over the last five
time bins (corresponding to roughly an ISCO period) in order to assess the degree of temporal variance of the original \mI calculations.
Here, we additionally address statistical variance arising from the limited number of grid cells and 
photon bundles in our \mc calculations. Figure \ref{fig:spec_variance} plots the spectra together
with the statistical variance over the same time, angle, and region bins used in the previous plots. The shaded regions
represent contours of 1$\sigma$ fluctuations for the S3Ea0 (top) and S3Ea9 (bottom) cases derived
from more than $10^7$ escaped bundle contributions.
These regions are qualitatively similar to the temporal variance plots, emphasizing
the high degree of convergence from thermal emissions, including disk (inner and bulk) as well as high-energy free-free
emissions from the optically thin funnel. The lack of extended shaded regions for case S3Ea0
is consistent with our observations that this model experienced very little Compton scattering.
The statistical variance is greater in hot, spinning black hole environments
at photon energies between $10^2$ - $10^3$ keV due to
the relatively small fraction of bundles being scattered to these energies.
Variance across all other frequencies is otherwise low, justifying our choice of grid resolution
and bundle sourcing rate.

\section{Emission maps}
\label{sec:radiationMaps}

\begin{figure}
\centering
\includegraphics[width=0.8\textwidth]{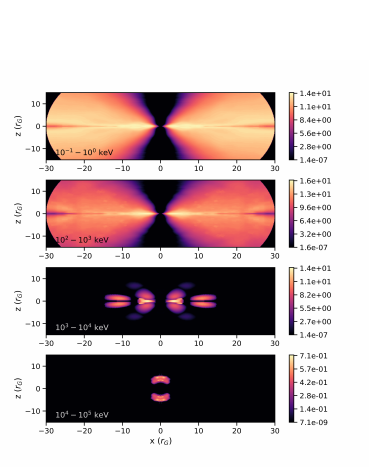}
\caption{
Spatial map plotting the logarithm of the average emission energy per volume (ergs/cm$^3$) in the birth zones of escaping
\mc bundles, with energies between 0.1 - 1 keV (top row), $10^2$ - $10^3$ keV (second row), $10^3$ - $10^4$ keV (third row), 
and $10^4$ - $10^5$ keV (bottom).
The maps are derived from model S3Ea9, azimuthally averaged, accumulated over all \mc cycles, and shown here at time $\sim 10^4 t_G$
after evolving for a period of 20 $t_{\text{ISCO}}$.
}
\label{fig:radmap_case6}
\end{figure}

Post-processing with Monte Carlo allows us to construct spatial radiation maps identifying the origins of
escaping photon bundles as a function of their detected frequencies (the frequency measured by a distant observer),
by accumulating (and storing) radiation energies emitted in the birth zones of each bundle when created.
These maps do not inform where photons scatter and acquire their observed energies, but
they do provide a map of where they were created.
Figure \ref{fig:radmap_case6} shows four panels displaying the escaping radiation map for case S3Ea9 representing different sources of emission:
soft thermal photons between 0.1 and 1 keV, originating in the bulk disk photosphere (top row); 
high temperature emissions between $10^2$ and $10^3$ keV,
originating from the inner $< 10 r_G$ regions of the disks (second row);
Compton scattered photons between $10^3$ and $10^4$ keV, also originating mostly from the very inner regions of the disks (third row);
and the highest free-free emissions, between $10^4$ and $10^5$ keV, created at the base of the funnel region along the pole (bottom row).

The maps are generated by accumulating over time the energy density (ergs/cm$^3$) sourced in cells
corresponding to the birth locations of escaping bundles, then binning them into 8 frequency groups
(one group per decade of energy). The sourced energy is additionally averaged azimuthally and
across the midplane to improve statistics before being projected onto a grid of $300\times360$ cells in the $x-z$ plane 
spanning a radius of $30 r_G$ with full polar extent,
corresponding to $0.2 r_G$ and $0.5^\circ$ resolutions in radius and angle, respectively.
To improve statistics even more, these maps are generated by extending the integration time to 20 $t_{\text{ISCO}}$,
roughly five times longer than the calculations presented in the previous sections.

We see that escaping photons in the lowest energy range (top panel) are emitted outside a conical shell
above and below the disk mid-plane, identifying the effective photosphere.
The midplane region below the photosphere is dark because photons are almost
entirely reabsorbed by the free-free process there.
Because Comptonization tends to dominate hot thermal emissions produced by the inner
(mostly relaxed) portion the disk, these source maps cannot distinguish between
thermal and Compton scattered photons. Apart from the magnitudes, maps across any decade band between 1 and $10^3$ keV are similar
and characterized by the map displayed in the second row.
Emissions between $10^3$ and $10^4$ keV (third row) represent regions contributing the highest energy photons 
upscattered by hot electrons. Collectively, the brightest regions in the top three panels therefore indicate where
thermally emitted photons can be upscattered efficiently, generally lying in the proximity of the photosphere and
black hole horizon (but outside the funnel) where gas temperatures and densities are high enough to affect inverse Comptonization.

Escaping high frequency bundles are produced in decreasingly smaller regions, generally
restricted to areas close to the horizon beyond the photosphere, transitioning from predominately disk to funnel emissions.
The bottom panel in Figure \ref{fig:radmap_case6} shows the very highest energy bundles created by free-free emissions in the hot,
but optically thin, funnel (aligned with the spin axis) with most of the luminosity being sourced at radii $\lesssim 5 r_G$.

\section{Model Comparisons}
\label{sec:comparison}

\begin{figure}
\centering
\includegraphics[width=0.6\textwidth]{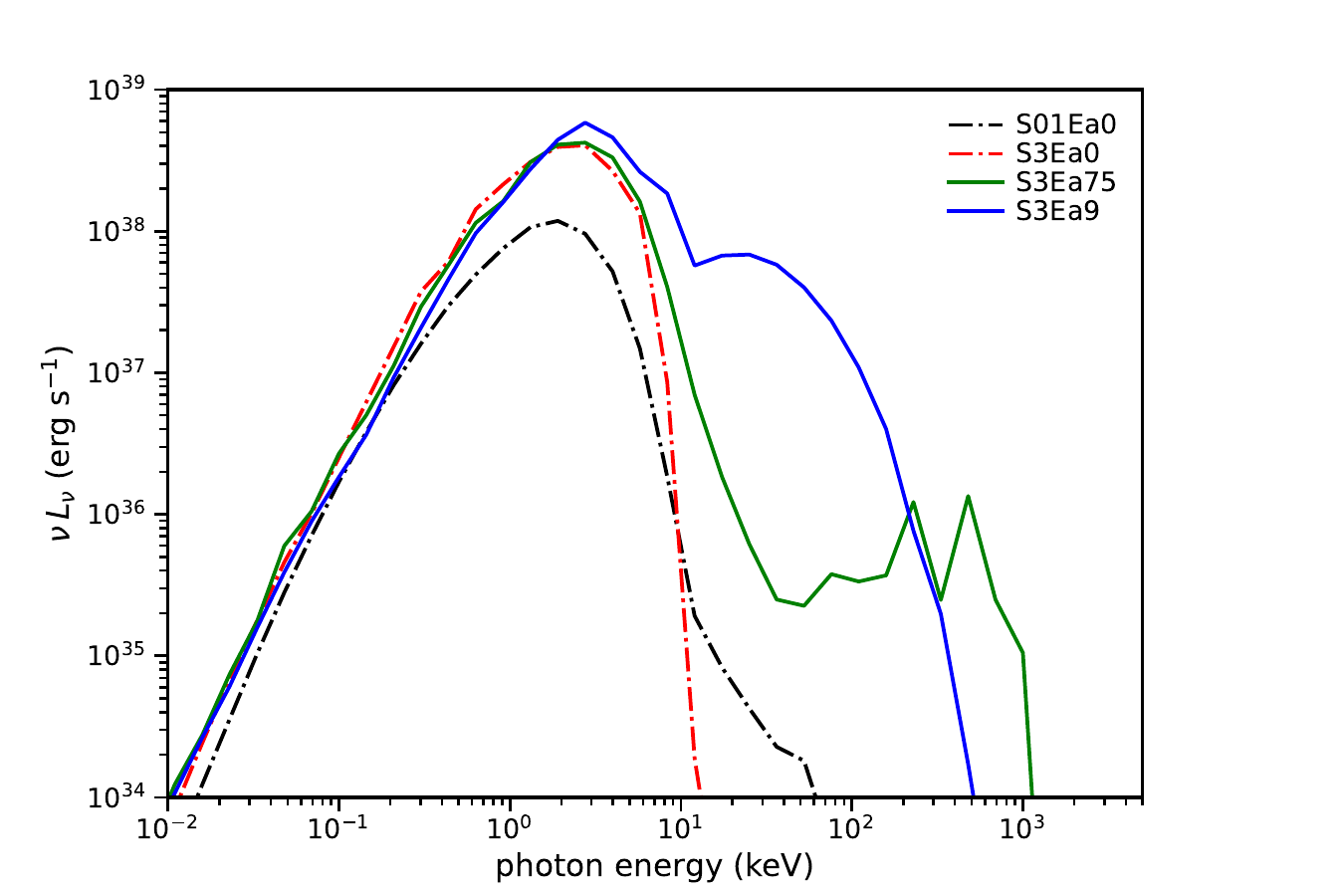}
\caption{
Comparison of the post-processed Monte Carlo spectra from all four cases zoomed in to highlight the similarities
in the dominant thermal peak at $\sim 2.5$ keV and the differences across the inverse Compton scattered emissions between
10 and $10^3$ keV. 
The spinning (non-spinning) black hole cases are represented with solid (broken) lines.
}
\label{fig:spectra_all}
\end{figure}

Figure \ref{fig:spectra_all} plots the \mc spectra from all four disk models, zoomed in to facilitate comparisons of
our simulated outputs to observations of black hole X-ray binaries (BHXRB).
All models share a similar low-energy, thermal profile peaking between 2 and $2.5$ keV, 
that extends into an increasingly Compton hardened component correlating with black hole spin.
As expected, they resemble observed spectra typical of BHXRBs in the soft to intermediate states, but with the highest
spin models appearing more closely aligned with the intermediate state.
In addition to the location of the common thermal peak, we observe three of the four models that produced
similar rates of mass accretion (S3Ea0, S3Ea75, and S3Ea9), agree also
in the magnitude of the outgoing luminosity, roughly $5\times10^{38}$ erg/s at 2.5 keV.
The exception is model S01Ea0 which produced significantly lower peak (by a factor of 4) and integrated luminosities due to
it being much thinner, predominantly gas-pressure supported, with less radiative cooling, and
supporting a lower mass accretion rate at large radii where most of the soft thermal emissions originate.

The zero spin models S01Ea0 and S3Ea0 are strongly dominated by thermal disk emissions
and match outputs, for example, from LMC X-3 in shape, peak frequency, and luminosity (for the case of S3Ea0) \citep{Gierlinski04, Davis06, Fragile23}.
Contrary to the results of \citet{Kinch21}, a comparison of these two simulations suggests a hardening of
the spectrum with increased accretion rate. However as we noted earlier, it is not entirely clear if hardening is attributed
to the accretion rate or to a global reconfiguration of the S3Ea0 disk due to the thermal instability.

As spin increases, our model spectra (S3Ea75 and S3Ea9) begin to deviate from the soft state into
a hardened intermediate state shaped primarily by the compact
coronal regions developing ever closer to the black hole with increasing spin.
Temperatures in these regions increase with spin due to the greater gravitational binding energy being released as
the event horizon shrinks and the inner edges of the disk reach closer to the black hole.
Because the hot peripheral regions of the disk thicken considerably while encroaching deeper into the funnel
toward the pole (compare the top and bottom rows of Figure \ref{fig:diskimages}), 
more escaping photons are subjected to hot scattering atmospheres,
contributing to greater Compton-hardened emissions above 10 keV.

Model S3Ea75, with moderate spin $a_*=0.75$, begins to show obvious signs of hardening from Compton
upscattering but nonetheless emits a thermally dominant X-ray spectrum
with high energy ($>10$ keV) features plateauing at roughly similar (albeit lower) relative magnitude 
compared to what is observed from LMC X-3. At the higher rate of spin represented by model S3Ea9, the shape
of the spectrum more resembles the intermediate
state of Cyg X-1 \citep{Zdziarski02, Gierlinski03}, rather than the soft state of LMC X-3.

Finally we note that the photon indices of models S01Ea0 and S3Ea75, evaluated between 5 and 100 keV,
are $\Gamma = 3.8$ and 3.6, respectively. Both fall within the range extracted from earlier numerical
models \citep{Kinch19} and observed BHXRBs in the thermal or steep power-law states \citep{McClintock06}.
Model S3Ea0 has a significantly steeper index of $\Gamma=6.5$, consistent with this model having developed
cooler coronal (and inner disk) regions severely limiting high-temperature emissions and Compton interactions beyond the soft
thermal profile (see column (b) of Figure \ref{fig:spec_cases1and2}). 
The index $\Gamma=2.0$ derived from model S3Ea9 is more typical of BHXRBs in the intermediate to hard states.

\section{Conclusions}
\label{sec:conclusion}

We have presented spectral X-ray outputs from the post-processing of four general relativistic, multi-frequency radiation MHD simulations 
of thin accretion disks with different mass accretion rates (0.5 to $3 c^2/L_\mathrm{Edd}$), black hole spins (0 to 0.9), and
luminosities (0.2 to $1.3 L_\mathrm{Edd}$). Because the disks are thin, they are challenging to
resolve computationally, and their infall times are longer than for thick disks, so they require especially long times to settle.
The unsettled nature of the disks leads to significant variances in spectral predictions, particularly at frequencies greater
than the soft thermal peak. These uncertainties manifest as temporal variations in X-ray emissions from  
both the original \mI transport simulations and the post-processed \mc results presented in this report.

Despite these difficulties, we are able to demonstrate a general agreement between the \mI (calculated inline)
and \mc (calculated by post-processing) results. The \mc calculations presented here are carried out over a broader energy range
(8 decades, compared to 6), and at higher spectral
fidelity (roughly four times more refined in energy), resolving in greater
detail the various contributing emissions, including the very high free-free features coming from coronae
developing near the black hole not covered by the \mI energy grid.
Both radiation solvers incorporate similar physics interactions, including fully general relativistic transport,
gravitational redshifts, Doppler shifts, free-free emission and absorption, and Compton scattering of photons with electrons.

Our studies elucidate essentially four emission sources contributing to the overall spectra,
common to all of our calculations: (1) the dominant soft component representing emissions off the photosphere
of the disk body peaking between 2 - $2.5$ keV,  
(2) direct emissions from the hot innermost regions of the disks contributing
to a relatively weaker radiation flux between 10 and 100 keV, (3) inverse Compton scattered
photons affecting the spectrum out to several hundred keV, and (4) free-free emissions
from hot, optically thin coronal regions near the black hole responsible for the most energetic X-rays ($>10^4$ keV).

Spectral emissions in most of our models resemble the soft, thermally dominant state with photon
indices (computed between 5 and 100 keV) greater than 3.5, as expected for thin disks,
but with a hardening trend towards the intermediate state with increasing black hole spin due to inverse Compton effects.
Over the same energy range, our highest spin model S3Ea9 is characterized by a photon index of $\Gamma=2$, typically associated with the hard state.

Unfortunately we were not able to investigate with certainty the effects of different accretion rates on the spectral outputs.
The two zero-spin cases initialized with different $\dot{m}_\mathrm{NT}$, S01Ea0 and S3Ea0, evolved on very different settling trajectories. 
S3Ea0 suffered significantly from the thermal instability, collapsing the disk to a much more compact configuration than expected.
We are thus unable to determine if spectral differences (a hardening of the spectrum between 10 and 100 keV with increasing accretion rate)
are attributed to the different accretion rates or to the reconfigured
state of the S3Ea0 disk. However it is interesting to note that S3Ea0 settled into a slightly
cooler disk with significantly cooler coronal regions, suppressing inverse Compton scattering almost entirely.

The two spinning cases considered here did not suffer the thermal instability, allowing us to comment conclusively on the effects
of black hole spin. We find a systematic (monotonic) hardening of the spectrum with increasing black hole spin, leading
to greater emissions at energies between 10 and several hundred keV with brighter integrated luminosities. 
This is supported by both the \mI and \mc methods, which agree nicely across the entire spectrum.
These enhanced emissions are generated by inverse Compton scattering as photons propagate through hot coronal gases
developing near the innermost regions of the disks, at temperatures that increase with increasing spin.
At near maximal rates of spin, inverse Compton scattering is capable of producing photon distributions
with spectral luminosities at $\sim100$ keV comparable to the soft thermal component.

\begin{acknowledgments}
This work was performed under the auspices of the U.S. Department of Energy 
by Lawrence Livermore National Laboratory under Contract DE-AC52-07NA27344. D. P. was supported by the U.S. Department of Energy, National Nuclear Security Administration, Minority Serving Institution Partnership Program, under Award DE-NA0003984. 
P.C.F. gratefully acknowledges support from the National Science Foundation under grant AST-1907850 and NASA through award No 23-ATP23-0100. 
\end{acknowledgments}

\vspace{5mm}
\facilities{LLNL (LC/CZ-Ruby), XSEDE (Stampede2)}

\software{Cosmos++ \citep{Anninos05,Anninos17,Fragile14,Fragile19,Anninos20,Roth22}}

\bibliographystyle{aasjournal}


\end{document}